\documentclass[12pt]{article}
\usepackage{epsfig}
\usepackage{amsfonts}
\usepackage{amsopn}
\usepackage{amsmath}
\usepackage{verbatim,amsthm}
\title
{\vskip -50 pt
\begin{flushright}
\normalsize\rm NORDITA-2019-97
\end{flushright}
\vskip 20 pt
 Brane mechanism of spontaneously generated gravity
}

\author{
 A. A. Zheltukhin 
  \thanks{E-mail: aaz@nordita.org} 
  \\
Kharkov Institute of Physics and Technology, \\
1, Akademicheskaya St., Kharkov, 61108, Ukraine \\  
NORDITA, Royal Institute of Technology and Stockholm University,\\
Roslagstullsbacken 23, SE-106 91 Stockholm, Sweden 
}

\date{}
\begin{document}

\maketitle

\begin{abstract}

The braneworld scenario is studied and the effective action of 3-brane living in 5-dim. Minkowski space is   constructed. This action is proved to be invariant under spontaneously broken scale and Weyl symmetries  and to encode a model of quadratic gravity generalizing the Starobinsky model. The symmetry breaking generates the Hilbert-Einstein term with the Newton constant $G_{N}\sim\frac{1}{\mu^{2}}$, where $\mu$ is a mass scale equal to the mean curvature of the vacuum hyper-worldsheet of 3-brane. This result proposes a brane modification of the mechanism of spontaneously generated gravity.

\end{abstract}

\bigskip

Observation data from Planck \cite{Planck} and other high precision experiments (see \cite{Lids} and refs. there) support the conception of inflation in cosmology. 
String/M-theory unifying QFT and gravity in 11-dimensional space-time are considered as a suitable basis for solution of this problem provided the extra dimensions are compactified to the Planck scale $l_{Pl}\sim G_{N}^{1/2}\sim 10^{-33}$. The original idea of Kaluza-Klein has supposed  that all fields in nature propagate in the same number of dimensions. Later this condition was weakened by the asumptions allowing trapping of matter or the existence of noncompact extra dimensions \cite{RS, GMZ, NW}. Studies in this direction resulted in interesting models solving the hierarchy problem between the weak and Planck scales, as well as explaining the observed weakness of the Newton gravity \cite{ADD, AADD, RaSu1, RaSu2}. Their result was that fields of the Standard Model might be confined on a 3-brane embedded into 5-dim. space-time, while the gravity propagated also in higher dimensions [10-15] \nocite{ShTy, HoWit, DShi, Sund, LOSW, DGP}. This confirmed the $1/r^{2}$ Newton law corrected by the term $\sim 1/r^{4}$ that might in principle be tested at distances of about 1 mm, remembering that gravity was accurately measured at distances of $\sim$ 1 cm. The fact  that extra dimensions can be much larger than the weak scale $\sim 10^{-16}$ makes 
the braneworld scenario promising for the unification of SM and gravity. This gives rise to the question about emergence of the Hilbert-Einstein gravity from a 3-brane identified with our Universe. Here we study the structure of non-linearities of 3-branes embedded into the 5-dim. Minkowski space on the basis of the Gauss-Cartan differential-geometric approach \cite{Car}. 
This approach treats strings and branes in terms of their  hyper-worldsheets (hyper-ws) embedded into higher-dimensional spaces [17-21] \nocite{RL, Omn, BN, Zgau2, Z_rmp}. We build an effective action for the 3-brane and reveal that this action encodes a model of quadratic gravity [22-29] \nocite{Strb, Smol, Zee, Adl, Frad, LPPS, Stel, Zwi} with the spontaneously broken Weyl and scale symmetries. This breaking creates the Hilbert-Einstein action with the Newton constant 
$G_{N}\sim\frac{1}{\mu^{2}}$ in correspondence with the mechanism of spontaneously generated gravity suggested in [23-25] \nocite{Smol, Zee, Adl}. The above constant $\mu$, equal to the  mean curvature of the vacuum 4-dim hyper-ws embedded into  $\bf{R^{1,4}}$, introduces a mass scale breaking the scale invariance.  

1. We explore effective  action for p-branes in terms of the metric $g_{\mu\nu}(\xi)$ 
and extrinsic curvature $l_{\mu\nu}(\xi)$ of their hyper-worldsheets $\Sigma_{p+1}$
 in the (p+2)-dim. Minkowski space $\bf{R^{1,p+1}}$   
\begin{eqnarray}\label{actncmc}
S_{p} =\frac{1}{k_{p}{}^{2}} \int d^{p+1}\xi\sqrt{|g|}
(\frac{1}{2}\nabla_{\mu}l_{\nu\rho}\nabla^{\{\mu}l^{\nu\}\rho}
-\nabla_{\mu}l^{\mu}_{\rho}\nabla_{\nu}l^{\nu\rho} + {\bf U}_{p}(g,l)).
\end{eqnarray}
The tensors $g_{\mu\nu}, l_{\mu\nu}$ 
define $\Sigma_{p+1}$ modulo its global translations and rotations.
 For p=3 and $\bf{R^{1,4}}$ we expect this action to be a generally covariant extension of massless $\phi^{4}$ theory which has only one dimensionless coupling constant $k_{3}$ depending on 3-brane tension (see \cite{Z_mpla}).  In this case the scalar potential ${\bf U}_{3}$ is represented by the quartic polynomial
\begin{eqnarray}\label{U}
{\bf U}_{3}=\frac{2}{3}TrlTp(l^{3}) - \frac{1}{2}(Tr(l^{2})^{2} + b_{2}Tr(l^{2})(Trl)^2 +  b_{4}(Trl)^{4} + b'_{4} Tr(l^{4}) + c_{3},
\end{eqnarray}
 where $b_{2}, b_{4}, b'_{4}$ are dimensionless parameters. We use  the canonical dimensions of the fields and coordinates in the system $\hbar=c=1$, i.e. $[\xi]=length, [l_{\mu\nu}]=(length)^{-1}$. 
The invariants of diffeomorphisms  $Tr(l^{n})$ are formed by covariant contractions of n symmetric
 tensors $l_{\mu\nu}$  
 \footnote{In our previous papers we used the German abbreviation SpF for the TrF.}  
\begin{eqnarray}\label{trace}
Trl=l_{\mu\nu}g^{\mu\nu}, \ \  Tr(l^{2})=l_{\mu\nu}l^{\nu\mu}, \ \ \ \ Tr(l^{3})=l_{\mu\rho}l^{\rho\gamma}l_{\gamma}{}^{\mu}, \ \ \ \  Tr(l^{4})=l_{\mu\rho}l^{\rho\gamma}l_{\gamma\lambda}l^{\lambda\mu}.
\end{eqnarray}
When p=3 and the cosmological constant $c_{3}=0$,  the action (\ref{actncmc}) with the potential (\ref{U}) becomes invariant under the global Weyl and scale transformations.

To prove this statement we choose the Weyl symmetry transformations for $g_{\mu\nu}$ as \begin{eqnarray}\label{weylg}  
 \xi^{\prime\mu}= \xi^{\mu}, \, \, \, g_{\mu\nu}^{\prime}(\xi^{\prime})= e^{2\alpha}g_{\mu\nu}(\xi). 
\end{eqnarray}
This gives the transformation law for the (p+1)-dim. differential world-volume 
\begin{eqnarray}\label{weylg'} 
d^{p+1}\xi^{\prime}\sqrt{|g^{\prime}(\xi^{\prime})|}
=e^{(p+1)\alpha}d^{p+1}\xi\sqrt{|g(\xi)|}.
\end{eqnarray}
The change of (\ref{weylg'}) must be compensated by transformations of the Lagrangian density in (\ref{actncmc}. Using Eq. (\ref{weylg'}) we find that the kinetic term prescribes the transformation for $l_{\mu\nu}$ 
\begin{eqnarray}\label{wkin}  
l_{\mu\nu}^{\prime}(\xi^{\prime})= e^{-\alpha[-3+ \frac{p+1}{2}]}l_{\mu\nu}(\xi). 
\end{eqnarray}
The potential term (\ref{U})(with $c_{p}=0$) dictates the Weyl transformation law  to be  
\begin{eqnarray}\label{wpot}  
l_{\mu\nu}^{\prime}(\xi^{\prime})= e^{-\alpha[-2+ \frac{p+1}{4}]}l_{\mu\nu}(\xi). 
\end{eqnarray}
The laws  (\ref{wkin}) and (\ref{wpot}) coincide for $p=3$. It proves the invariance of the effective 
action  $S_{3}$  (\ref{actncmc}) with $c_{3}=0$ under the global Weyl transformations 
\begin{eqnarray}\label{wtot} 
 \xi^{\prime}{}^{\mu}= \xi^{\mu}, \, \, \, \,  \, g_{\mu\nu}^{\prime}(\xi^{\prime})= 
 e^{2\alpha}g_{\mu\nu}(\xi),  \, \, \, \,  \, l_{\mu\nu}^{\prime}(\xi^{\prime})=e^{\alpha}l_{\mu\nu}(\xi).
\end{eqnarray}
The action (\ref{actncmc}) with $c_{3}=0$ is also invariant under the global dilatation symmetry.
 
In field theory a field $\varphi(x)$ of the dimension $q$ has the following scaling transformations 
\begin{eqnarray}\label{fdil} 
 x^{\prime}{}^{m}= e^{-\lambda}x^{m}, \, \, \, \,  \, \varphi^{\prime}(x^{\prime})=e^{q\lambda}\varphi(x)  
\end{eqnarray}
 where $\lambda$ is a rescaling parameter. 
Using these rules we find the form variation $\bar{\delta}\varphi(x)$ of a field $\varphi$ under infinitesimal transformations 
\begin{eqnarray}\label{infdil} 
 \bar{\delta}\varphi(x)= \delta\lambda[q + x^{m}\partial_{m}]\varphi(x), \, \, \, \, \,
 \delta\varphi(x)=q\varphi(x)\delta\lambda.
\end{eqnarray}
Keeping in mind that  $g_{\mu\nu}(\xi)$ is dimensionless we have 
 $\bar{\delta}g_{\mu\nu}(\xi)= \delta\lambda\xi^{\rho}\partial_{\rho}g_{\mu\nu}(\xi)$ that gives   
\begin{eqnarray}\label{gdil} 
\xi^{\prime}{}^{\mu}= e^{-\lambda}\xi^{\mu}, \, \, \,   g^{\prime}_{\mu\nu}(\xi^{\prime})=g_{\mu\nu}(\xi) \,  \longrightarrow \, \,  d^{p+1}\xi^{\prime}\sqrt{|g{\prime}(\xi^{\prime})|}=e^{-(p+1)\lambda}d^{p+1}\xi\sqrt{|g(\xi)|},
 \end{eqnarray} respectively.  
Eqs. (\ref{gdil}) prescribe the following  transformation rule for a scalar potential ${\bf U}_{p}$ 
\begin{eqnarray}\label{Vdil} 
{\bf U}_{p}^{\prime}(g^{\prime}_{\mu\nu}(\xi^{\prime}), l^{\prime}_{\mu\nu}(\xi^{\prime}))
= e^{(p+1)\lambda}
{\bf U}_{p}(g_{\mu\nu}(\xi), l_{\mu\nu}(\xi)).
\end{eqnarray}
As a result,  we obtain the scaling transformation law for $l_{\mu\nu}(\xi)$
\begin{eqnarray}\label{lpotdil}  
l_{\mu\nu}^{\prime}(\xi^{\prime})= e^{\lambda\frac{(p+1)}{4}}l_{\mu\nu}(\xi). 
\end{eqnarray}
To find the law for $l_{\mu\nu}(\xi)$ prescribed by the kinetic part of the action (\ref{actncmc}) note that  
 \begin{eqnarray}\label{criscdil} 
\partial_{\mu}{}^{\prime}=e^{\lambda}\partial_{\mu}, \, \, \Gamma^{\prime}_{\mu,\, \nu \rho}(\xi^{\prime})=e^{\lambda}\Gamma_{\mu,\, \nu \rho}(\xi) \longrightarrow \nabla^{\prime \mu}=e^{\lambda}\nabla^{\mu},
\end{eqnarray}
where $\Gamma_{\mu,\, \nu \rho}(\xi)$ are the Christoffel symbols for  $\Sigma_{p+1)}$. 
It yields the following law for $l_{\mu\nu}(\xi)$ 
\begin{eqnarray}\label{lkindil}  
l_{\mu\nu}^{\prime}(\xi^{\prime})= e^{\lambda\frac{(p-1)}{2}}l_{\mu\nu}(\xi). 
\end{eqnarray} 
The transformation rules (\ref{lpotdil}) and (\ref{lkindil}) coincide only if $p=3$. 
This prows invariance of $S_{3}$ (\ref{actncmc}) under the global scaling transformations
\begin{eqnarray}\label{diltot}  \,
 \xi^{\prime}{}^{\mu}= e^{-\lambda}\xi^{\mu}, \, \, \, \, \, g_{\mu\nu}^{\prime}(\xi^{\prime})= 
 g_{\mu\nu}(\xi),  \, \, \, \,  \, l_{\mu\nu}^{\prime}(\xi^{\prime})= e^{\lambda}l_{\mu\nu}(\xi).
\end{eqnarray}
Thus, $S_{3}$ (\ref{actncmc}) with $c_{3}=0$ turns out to be invariant under the $U(1)\times U(1)$ abelian symmetry 
\begin{eqnarray}\label{w-dil}  \,
 \xi^{\prime}{}^{\mu}= e^{-\lambda}\xi^{\mu}, \, \, \, \, \, g_{\mu\nu}^{\prime}(\xi^{\prime})= 
 e^{2\alpha}g_{\mu\nu}(\xi),  \, \, \, \,  \, l_{\mu\nu}^{\prime}(\xi^{\prime})
 =e^{\alpha +\lambda}l_{\mu\nu}(\xi).
\end{eqnarray}
Eqs. (\ref{w-dil}) show that the abelian subgroup $U_{+}$  of $U(1)\times U(1)$ formed by $\alpha=\lambda$ 
\begin{eqnarray}\label{U1} 
 \xi^{\prime}{}^{\mu}= e^{-\lambda}\xi^{\mu}, \, \, \, \, \, g_{\mu\nu}^{\prime}(\xi^{\prime})= 
 e^{2\lambda}g_{\mu\nu}(\xi),  \, \, \, \,  \, l_{\mu\nu}^{\prime}(\xi^{\prime})
 =e^{2\lambda}l_{\mu\nu}(\xi)
\end{eqnarray}
belongs to the group of diffeomorphisms of $\Sigma_{4}$ and it protects $U_{+}$ from breakdown.
 It means that breaking of the Weyl symmetry demands breaking of the scale symmetry of $S_{3}$
 (\ref{actncmc}) with  $c_{3}=0$. This observation hints that the scale symmetry can be broken 
without breaking of the  diffeomorphism symmetry by imposing the diff invariant condition 
\begin{eqnarray}\label{U2mcc} 
 Trl=\mu, \, \, \, \, \, \mu=const\neq0
\end{eqnarray}
which also breaks the  Weyl symmetry, but preserves the diff symmetry in view of  
\begin{eqnarray}\label{scalbr} 
 Trl^{\prime}(\xi^{\prime})=e^{\lambda}Trl(\xi),  \, \, \, \, \, \, \, \,
  Trl^{\prime}(\xi^{\prime})=e^{-\alpha}Trl(\xi).
\end{eqnarray}
The condition (\ref{U2mcc}) creates the new constant $\mu$ with the dimension 
$[\mu]=[l_{\mu\nu}]=(length)^{-1}$ in addition to $k_{3}$ from (\ref{actncmc}). 
So, the condition (\ref{U2mcc}) destroys the Weyl and scale symmetries of $S_{3}$, as well as its discrete $Z_{2}$ symmetry
 \begin{eqnarray}\label{discr} 
 l_{\mu\nu}^{\prime}(\xi^{\prime})=-l_{\mu\nu}(\xi).
\end{eqnarray}
 Below we show that all these symmetries are spontaneously broken, because the condition (\ref{U2mcc})  realizes the extremal of the potential ${\bf U}_{3}$.

2. The equation of motion for $l_{\mu\nu}$ following from $S_{3}$ (\ref{actncmc}) is 
\begin{eqnarray}\label{eqrl}
\frac{1}{2}\nabla_{\mu}\nabla^{[\mu}l^{\{\nu]\rho\}}
=-[\nabla^{\mu},  \nabla^{\{\nu}] l_{\mu}{}^{\rho\}}
+ \frac{\partial {\bf U}_{3}}{\partial l_{\nu\rho}},   
\end{eqnarray}
whereas the evolution of $g_{\mu\nu}$ is defined by the Gauss embedding conditions for $\Sigma_{4}$
\begin{eqnarray}\label{cRcd}
R_{\mu\nu\gamma\lambda}(g)=l_{\mu\gamma} l_{\nu\lambda} - l_{\nu\gamma} l_{\mu\lambda}.
\end{eqnarray}
The latter combined with the Bianchi identities 
\begin{eqnarray}\label{BI}
[\nabla_{\mu} , \, \nabla_{\nu}] l^{\gamma\rho}
=R_{\mu\nu}{}^{\gamma}{}_{\lambda} l^{\lambda\rho}  
+ R_{\mu\nu}{}^{\rho}{}_{\lambda} l^{\gamma\lambda}.   
\end{eqnarray}
permits to write the commutator in the r.h.s. of (\ref{eqrl}) in the form 
\begin{eqnarray}\label{BIsim}
\frac{1}{2}[\nabla^{\mu} , \, \nabla^{\{\nu}]l_{\mu}^{\rho\}}
=(l^{2})^{\nu\rho} Trl - l^{\nu\rho} Tr(l^2),
\end{eqnarray}
where $Tr(l^{2}):=l_{\mu\rho}l^{\rho}_{\nu}g^{\mu\nu}$.
As a result,  Eq. (\ref{eqrl}) transforms to the PDE for $l_{\mu\nu}$ 
\begin{eqnarray}\label{eqr1B}
-\frac{1}{4}\nabla_{\mu}\nabla^{[\mu}l^{\{\nu]\rho\}}
=(l^{2})^{\nu\rho}Trl - l^{\nu\rho}Tr(l^{2})- \frac{1}{2}\frac{\partial {\bf U}_{3}}{\partial l_{\nu\rho}}. 
\end{eqnarray}
To clarify our approach we consider a simple potential ${\bf U}_{3}$ with zero phenomenological constants
$b_{2}=b_{4}=b'_{4} =0$ changing  ${\bf U}_{3}\rightarrow U$ with $U$ given by 
\begin{eqnarray}\label{redU}
 U =\frac{2}{3}TrlTr(l^{3}) - \frac{1}{2}(Tr(l^{2})^{2}.
\end{eqnarray}
Then EOM (\ref{eqr1B}) simplifies to the PDE
\begin{eqnarray}\label{eqr1m}
\frac{1}{2}\nabla_{\mu}\nabla^{[\mu}l^{\{\nu]\rho\}} 
= \frac{2}{3} Tr(l^{3})\frac{\partial {Trl}}{\partial l_{\nu\rho}}
\end{eqnarray}
which is the Euler-Lagrange equation generated by the Weyl and scale invariant action 
\begin{eqnarray}\label{actncmsf}
S =\frac{1}{k_{3}^{2}}\int d^{4}\xi\sqrt{|g|}
(\frac{1}{2}\nabla_{\mu}l_{\nu\rho}\nabla^{\{\mu}l^{\nu\}\rho} 
-\nabla_{\mu}l^{\mu}_{\rho}\nabla_{\nu}l^{\nu\rho}   
+\frac{2}{3}TrlTr(l^{3}) - \frac{1}{2}(Tr(l^{2})^{2}).  
\end{eqnarray}
An extremal $l^{\nu\rho}_{o}$ of the reduced potentail $U$ (\ref{redU}) is found from the equation 
\begin{eqnarray}\label{extrp-c}
(l^{2}_{o})^{\nu\rho}Trl_{o} -l^{\nu\rho}_{o}Tr(l^{2}_{o})=0
\end{eqnarray}
which can be written as
\begin{eqnarray}\label{extrp-c1}
l^{\nu}_{o\alpha}( l^{\alpha\rho}_{o}Trl_{o} - g^{\alpha\rho} Tr(l^{2}_{o}))=0.       
\end{eqnarray}
An evident solution of Eq. (\ref{extrp-c1}) is $l_{o\mu\nu}^{Mink.}=0$ which corresponds to usual 4-dim. Minkowski vacuum with $g_{o\mu\nu}=\eta_{\mu\nu}$. 
Vacuum breaks the Weyl symmetry simultaneously with the diff symmetry but it preserves the scale symmetry.  

More appealing extremal given by a non-degenerate matrix $l^{\nu}_{o\alpha}\equiv g^{\nu\gamma}l_{o\gamma\alpha}$ is 
\begin{eqnarray}\label{solextrp-c}
l_{o\mu\nu}= \frac{Trl_{o}}{4}g_{\mu\nu}, 
\ \ \ \ 
det\, l^{\mu}_{o\nu} \neq0.   
\end{eqnarray}
The solution  (\ref{solextrp-c}) yields the relations
 \begin{eqnarray}\label{degr}
(l_{o}^{n})_{\mu\nu}=(\frac{Trl_{o}}{4})^{n} g_{\mu\nu} \ \  \rightarrow \ \
Tr(l_{o}^{n})=4(\frac{Trl_{o}}{4})^{n}.   
\end{eqnarray}
The vacuum (\ref{solextrp-c}) is infinitely degenerate (similarly to the vacuum of (anti)ferromagnetic \cite{VZB} defined by the magnetization direction of its) being a linear representation 
of the symmetry groups of the action (\ref{actncmsf}). 
However, degeneracy with respect to the Weyl and scale groups is removed by the requirement
 that the extremal (\ref{solextrp-c}) satisfy EOM (\ref{eqr1m}). Here we notice that the l.h.s. of this equation  
 vanishes  
 if the following invariant vacuum condition is imposed 
\begin{eqnarray}\label{ccrvac} 
\nabla_{[\mu}l_{o\nu]\rho}=0  \ \ \ \longrightarrow  \ \ \ \nabla^{\rho}l_{o\rho\nu}=\nabla_{\nu}Trl_{o}\equiv\partial_{\nu}Trl_{o}.
 \end{eqnarray}
 The substitution of the extremal $l_{o\mu\nu}$ (\ref{solextrp-c}) in 
(\ref{ccrvac}) results in the condition 
\begin{eqnarray}\label{PCvac} 
\frac{1}{4}\partial_{\nu}Trl_{o}= \partial_{\nu}Trl_{o}.
 \end{eqnarray}
which requires   $Trl_{o}=constant$. So, we arrive at the earlier discussed 
condition (\ref{U2mcc}): $Trl_{o}=\mu$. The latter condition selects the desired extremal from the set (\ref{solextrp-c}) 
\begin{eqnarray}\label{vac}
l_{o\mu\nu}= \frac{\mu}{4}g_{\mu\nu}. 
\end{eqnarray}
The vacuum  (\ref{vac}) spontaneously breaks the Weyl, scale and $Z_{2}$ symmetries of 
the effective action (\ref{actncmsf}) introducing a mass scale $\mu$.
Noticing that $Trl$ is invariant under diffeomorphisms one can identify the trace 
with a massless scalar field  $\phi$ having a non-zero vev $<\phi>_{0}$ 
\begin{eqnarray}\label{dilat}
\phi:=Trl \ \ \  \rightarrow \ \ \     \phi_{0}:=<\phi>_{0}= Trl_{0}=\mu.   
\end{eqnarray}
One can handle $\phi$ as dilaton similarly to the proposal \cite{RV}, where a scale-invariant model of  quadratic gravity, including a scalar field coupled with gravity, was studied.

The extremal (\ref{vac}) of the effective field theory (\ref{actncmsf}) is associated  with a 3-brane,
 because it satisfies the Peterson-Codazzi embedding equation 
 \begin{eqnarray}\label{ccr} 
\nabla_{[\mu}l_{\nu]\rho}=0  
 \end{eqnarray}
(since $\nabla_{\rho}g_{\mu\nu}=0 $) for a hyper-ws  with codim 1.  
Eq. (\ref{ccr}) supplements Eq. (\ref{cRcd}) together with which they select the
 3-brane sector in the space of solutions of (\ref{eqr1m}). 
 So, the explicit solution  (\ref{vac}) permits to restore the  
 vacuum hyper-ws $\Sigma^{o}_{4}$ using the Gauss theorem  (\ref{cRcd})  
\begin{eqnarray}\label{cRcdV}
R_{o\mu\nu\gamma\lambda}=l_{o\mu\gamma}l_{o\nu\lambda} - l_{o\nu\gamma} l_{o\mu\lambda}.
\end{eqnarray}
Then the substitution of (\ref{vac}) in (\ref{cRcdV}) yields the vacuum Riemann tensor 
\begin{eqnarray}\label{extrRimf}
R_{o\mu\nu\gamma\lambda}
=(\frac{\mu}{4})^{2}(g_{o\mu\gamma}g_{o\nu\lambda}-g_{o\nu\gamma} g_{o\mu\lambda})
\end{eqnarray} 
which shows that $\Sigma^{o}_{4}$ is the space of constant curvature 
\begin{eqnarray}\label{vcurva4}
R_{o\mu\nu}=\frac{3}{(4)^{2}}\mu^{2} g_{o\mu\nu}, \ \ \ \ \ R_{o}:=g^{\mu\nu}R_{o\mu\nu}=\frac{3}{4}\mu^{2}. 
\end{eqnarray}
From (\ref{cRcdV}) we find the  Ricci tensor $R_{o\mu\nu}$ and 
the scalar curvature $R_{o}$ of $\Sigma^{o}_{4}$ 
\begin{eqnarray}
R_{o\nu\lambda}=-(l_{o}^{2})_{\nu\lambda} + \mu l_{o\nu\lambda}, \ \ R_{o}=-Sp(l_{o}^{2})+ \mu^{2},   \label{cub}  \\  
R_{o\nu\lambda}l_{o}^{\lambda\nu}= -Sp(l_{o}^{3}) + \mu Sp(l_{o}^{2}). \ \ \ \ \ \ \ \ \ \ \ \ \ \ \  \nonumber  
\end{eqnarray}      
Potential $U$ (\ref{redU}) on the extremal $l_{o\mu\nu}$ (\ref{vac}) takes the form
\begin{eqnarray}\label{redUvac}
 U_{0}:=U|_{l=l_{0}}=\frac{2}{3}\mu Tr(l_{o}^{3}) - \frac{1}{2}(Tr(l_{o}^{2})^{2}.
\end{eqnarray}
Relations (\ref{degr}) give  $U_{0}=\mu^{4}/96$.
 Next we observe that (\ref{redUvac}) can be written as 
\begin{eqnarray}\label{cub'}
U_{0}= -\frac{1}{2}R_{o}^{2} + \frac{\phi_{0}^{2}}{3}R_{o} 
- \frac{2}{3}R_{o\nu\lambda}l_{o}^{\lambda\nu}\phi_{0} + \frac{\phi_{0}^{4}}{6}. 
\end{eqnarray}
Using (\ref{cub'}) and vanishing of the kinetic terms in (\ref{actncmsf}) on extremal (\ref{vac})
 we obtain 
\begin{eqnarray}\label{actnvac}
S_{o} =\frac{1}{k_{3}^{2}}\int d^{4}\xi\sqrt{|g_{o}|}
(-\frac{1}{2}R_{o}^{2} + \frac{\phi_{0}^{2}}{3}R_{o} 
- \frac{2}{3}R_{o\nu\lambda}l_{o}^{\lambda\nu}\phi_{0} + \frac{\phi_{0}^{4}}{6}).
\end{eqnarray}
This shows  that $S_{o}$ is the first  term in the expansion of the  quadratic gravity action 
\begin{eqnarray}\label{actind}
S =\frac{1}{k_{3}^{2}}\int d^{4}\xi\sqrt{|g|}
(\frac{1}{2}\nabla_{\mu}l_{\nu\rho}\nabla^{\{\mu}l^{\nu\}\rho} 
-\nabla_{\mu}l^{\mu}_{\rho}\nabla_{\nu}l^{\nu\rho} \label{actngrav} 
\\
- \frac{1}{2}R^{2} + \frac{\phi^{2}}{3}R 
- \frac{2}{3}R_{\nu\lambda}l^{\lambda\nu}\phi + \frac{\phi^{4}}{6})  \nonumber
\end{eqnarray} 
around the extremal of the potential $U$ (\ref{redU}). In (\ref{actngrav}) we represent $l_{\mu\nu}$ as the sum 
\begin{eqnarray}\label{expal}
l_{\mu\nu}={\bar l}_{\mu\nu} + \frac{1}{4}\phi g_{\mu\nu}, \ \ \ \ \ \ \ \  Tr{\bar l}=0,
\end{eqnarray}
where ${\bar l}_{\mu\nu}$ is the traceless part of $l_{\mu\nu}$, i.e. $g^{\mu\nu}{\bar l}_{\mu\nu}=0$, 
in correspondence with (\ref{dilat}).
 Thus, the  effective scale invariant action  (\ref{actncmsf}) of 3-brane realizes the mechanism of spontaneously generated gravity. The resulting Newton constant 
 $G_{N}\sim\frac{1}{k_{3}^{2}\mu^{2}}$ is defined by the vev $\phi_{0}$ equal to the mean \bigskip 
curvature $\mu$ of the vacuum hyper-ws $\Sigma^{o}_{4}$.

\noindent{\bf Summary}
 
In the braneworld scenario we treat the action  (\ref{actind}) as an action of our universe. This gives
 an example of building new scale-invariant $R^{2}$ models in which a Brans-Dicke-like scalar $ \varphi$
is changed by tensors similar to $l_{\mu\nu}$.  In new models scalars like $\varphi$ will emerge in the form of composite fields similar to diff invariants $Trl\equiv g^{\mu\nu}l_{\mu\nu}$ with a non-zero vev $<Trl>_{0}$ introducing a new mass scale similar to $\mu$.
 The fact that scale-invariant models with scalars well describe inflation, reheating and fit the modern observations stimulates their extensions which introduce $l_{\mu\nu}$ for analyzing the current experiments in cosmology.

\noindent{\bf Acknowledgments}

The author is grateful to NORDITA and Physics Department 
of Stockholm University for kind hospitality and support.
These results were reported at the International Bogolyubov Conference
"Problems of Theoretical and Mathematical Physics"
(Moscow-Dubna, Sept. 9-13, 2019).
It is a pleasure to thank BLTP JINR, Dubna for warm hospitality and aid.
I am also grateful to R.B. Nevzorov, O.V. Teryaev,	H. von Zur-M{\"u}hlen, F. Wilczek, A. Yung, V.I. Zakharov and A. Zee for helpful discussions and comments.

\end{document}